\documentclass[usenatbib]{mnras}

\usepackage{graphicx}
\usepackage{amssymb,amsmath}
\usepackage{natbib}
\usepackage{hyperref}
\usepackage{cleveref}

\newcommand{\elik}{{\mathbf K}}
\newcommand{\elie}{{\mathbf E}}

\newcommand{\deltao}{\Delta_0}

\newcommand{\ko}{{k_0}}
\newcommand{\kpo}{{k'_0}}

\newcommand{\rc}{R_c}

\newcommand{\rf}{{R_{\rm p}}}

\newcommand{\kp}{{k'}}

\newcommand{\psiref}{\Psi_{\rm ref}}

\newcommand{\massloop}{M_{\rm loop}}
\newcommand{\massshell}{M_{\rm shell}}
\newcommand{\masssolid}{M_{\rm solid}}

\newcommand{\newkappa}{\kappa}
\newcommand{\newkappao}{\kappa_0}

\begin{document}

\title[The gravitational potential of toroids]{The exterior gravitational potential of toroids}
\author[J.-M. Hur\'e et al.]
{J.-M. Hur\'e$^{1}$\thanks{E-mail:jean-marc.hure@u-bordeaux.fr},
  B. Basillais$^{1}$,
  V. Karas$^{2}$,
  A. Trova$^{3}$,
  and   O. Semer\'ak$^{4}$\\
$^{1}$Laboratoire d'Astrophysique de Bordeaux, Univ. Bordeaux, CNRS, B18N, all\'ee Geoffroy Saint-Hilaire, 33615 Pessac, France\\
$^{2}$Astronomical Institute, Academy of Sciences, Bo\v{c}n\'{\i} II 1401, CZ-14100 Prague, Czech Republic\\
$^{3}$University of Bremen, Center of Applied Space Technology and Microgravity (ZARM), 28359 Bremen, Germany\\
$^{4}$Institute of Theoretical Physics, Faculty of Mathematics and Physics, Charles University, CZ-180 00 Prague, Czech Republic
}

\date{Received ??? / Accepted ???}
 
\pagerange{\pageref{firstpage}--\pageref{lastpage}} \pubyear{???}

\maketitle

\label{firstpage}

\begin{abstract}
We perform a bivariate Taylor expansion of the axisymmetric Green function in order to determine the exterior potential of a static thin toroidal shell having a circular section, as given by the Laplace equation. This expansion, performed at the centre of the section, consists in an infinite series in the powers of the minor-to-major radius ratio $e$ of the shell. It is appropriate for a solid, homogeneous torus, as well as for inhomogeneous bodies (the case of a core stratification is considered). We show that the leading term is identical to the potential of a loop having the same main radius and the same mass | this ``similarity'' is shown to hold in the ${\cal O}(e^2)$ order. The series converges very well, especially close to the surface  of the toroid where the average relative precision is $\sim 10^{-3}$ for $e\! = \!0.1$ at order zero, and as low as a few $10^{-6}$ at second order. The Laplace equation is satisfied {\em exactly} in every order, so no extra density is induced by truncation. The gravitational acceleration, important in dynamical studies, is reproduced with the same accuracy. The technique also applies to the magnetic potential and field generated by azimuthal currents as met in terrestrial and astrophysical plasmas.
\end{abstract}


\section{Introduction}

The derivation of reliable and compact expressions for the gravitational potential of massive toroids is a longstanding problem of dynamical astronomy, from planetary to galactic scales. This is essential not only  to examine the motion of test-particles and fluids orbiting around, in the classical framework as well as in general relativity \citep{ni05,subrkaras05,sesu10,tresaco11,iorio12}, but also to understand the conditions for the formation, evolution and stability of toroids themselves \citep{dyson1893b,hachisu86,chandra87,th90,wst92,sto93,chris93,hashi93,eri93,nier94,pdl97,horedttextbook2004,lss19}. While it is relatively easy to deduce the mass density corresponding to a given potential \citep[e.g.][]{binneytremaine87}, the inverse procedure is very complicated by analytical means, and it is almost impossible to go beyond the classical series representations and to get closed forms \citep{clement74,cohl01,petroff08}. Fully numerical approaches may be preferred for their apparent simplicity, but the computing times are generally large, often prohibitive at high spatial resolution, especially for very inhomogeneous configurations and/or very extended systems like discs. The numerical accuracy of discretization schemes is mainly limited when treating thin sources (having less than three spatial dimensions) whose field typically suffers a certain irregularity at their position.

In axial symmetry, the Green function ${\cal G}(\vec{r}|\vec{r}')$ of the Poisson equation involves the complete elliptic integral of the first kind $\elik$ whose argument (or modulus) gathers all the pertinent variables \citep{kellogg29, durand64,fu16}. The presence of a special function is a real obstacle when it is to be convolved with any non-trivial mass density $\rho(\vec{r}')$. One can overcome this difficulty by expanding $\elik$ over the modulus, but the ``dual'' nature of the series -- different for large and for short separations -- means piece-wise approximations whose connection requires technical efforts. This is done for instance in \cite{ba11} who match together the internal and the external potentials of the solid (i.e. homogeneous) torus from a minimization procedure.

This article brings a new contribution to this general and challenging problem. It is inspired by \cite{htkl19} who derived a reliable approximation for the {\em interior} potential of a toroidal shell of circular cross-section, based on a bivariate expansion performed at the pole (or focal ring) of the toroidal coordinates. At this singular point, all the partial derivatives of the Green function are exceptionally analytical. Unfortunately, the ``pole'' method does not apply outside the shell cavity because the line segment linking the focal ring to any exterior point crosses the shell where the Green function is basically singular. We generate accurate approximations for the {\em exterior} solution of the toroidal shell by expanding the axisymmetric Green function as a Taylor series {\it before integrating over the source}. As for the classical multipole expansion, the shell potential writes as an infinite series \citep[e.g.][]{mm18}, but our approach differs in that the origin of coordinates does not play a special role: the expansion is performed at the centre of the toroid section. Such an approach has been reported very recently by \cite{kondratyev18} in the case of the solid torus. The author writes the external potential in the form $\sum_n{\phi_n e^{2n}}$ ($e$ is the minor-to-major radius ratio, i.e. the torus parameter). He then uses the second-order expression to set constraints on the masses of thin, virialized rings orbiting an asteroid.\\

In this article, we go beyond the hypothesis made in \cite{ba11} and in \cite{kondratyev18} by considering inhomogenous systems too. In particular, we show that, when the toroid is radially stratified from the centre to the surface, only moments of the density need to be calculated. The method has an unexpected efficiency, not only at large distances, but also quite close to the surface of the toroid. The leading term has a correct behaviour at infinity as well as on the $Z$-axis, and it obeys the Laplace equation. As a matter of fact, these desirable properties are shared by {\em all} terms of the expansion. There is thus no spurious noise or extra density induced in space, whatever the truncation order. We treat orders $0$ to $2$ explicitly (a driver F90-program is appended). The resulting shell potential can be recast in the form of a ``modified monopole'' or in the form of an ``equivalent loop'', which concept has been discussed in \cite{sta83a}, while proofs are found in \cite{ba11} and \cite{kondratyev18}. We show that the exterior potential of the solid torus, which is of more astrophysical relevance than the shell, is easily deduced, with all the properties observed for the shell maintained. The method also applies to the determination of the vector potential and magnetic field of electromagnetism for toroids carrying a purely azimuthal current \citep{audrey18}.

The paper is organized as follows. In Sect. \ref{sec:potatpole}, the expression for the potential of a toroidal shell is given in its integral form. The axisymmetric Green function is expanded, in a bivariate manner, in Sect. \ref{sec:expansion}. The leading term is calculated and compared with the potential of a monopole (i.e. a point mass) and of a circular loop. Its precision is checked against an ``exact'' numerical reference in Sect. \ref{sec:efficiency}. The $1$st-order and $2$nd-order approximations are treated then similarly in Sect. \ref{sec:orders12}. In Sect. \ref{eq:connection} we show how the exterior and interior solutions match together at the shell surface. The procedure leading to the $n$th-order term is detailed in Sect. \ref{sec:gen}. The case of a solid torus is treated in Sect. \ref{sec:solidtorus}, while the case of core-stratified toroids is the aim of Sect. \ref{sec:inhomogeneous}. The formula for the gravitational acceleration is derived in Sect. \ref{sec:acceleration}. In particular, we show that the vertical component that rules hydrostatic equilibrium differs by a factor of about $2$ from Paczy\'nski's estimate valid for thin discs \citep{pacz78}.  This result is suited to examining the stability of rings \citep[e.g.][]{wt88}. From the radial component, we deduce the circular velocity of test particles orbiting in the equatorial plane. This formula can be helpful in explaining the deviations to the Kepler's law in massive systems \citep[e.g.][]{gds98}. Section \ref{sec:ab} is devoted to the magnetic potential due to toroidal currents (the leading term is derived). Two general comments are found in Sect. \ref{sec:comments}. Conclusions and perspectives are found in the last section.

\begin{figure}
\includegraphics[width=8.5cm,bb=170 140 600 400,clip=true,angle=0]{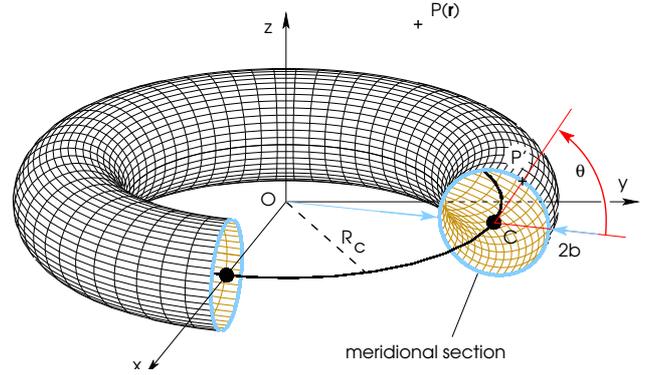}
\caption{The infinitely thin, toroidal shell (main centre O and main radius $\rc$) with a circular meridional section (centre C and core radius $b$).}
\label{fig:tshell.eps}
\end{figure}

\section{Potential of the toroidal shell}
\label{sec:potatpole}

We consider the simplest possible toroidal shell, as depicted in Fig. \ref{fig:tshell.eps}. The major radius is $\rc$ and the meridional section is circular, with centre C and minor radius
\begin{flalign}
b\equiv eR_c \le \rc,
\end{flalign}
where $e \in [0,1]$ denotes the shell parameter. We work in cylindrical coordinates $(R,Z)$, using the symmetry axis of the shell as the $Z$-axis, and $x$O$y$ as the plane of symmetry. For this specific problem, the Green function of the Poisson equation \citep[e.g.][]{kellogg29,durand64} is
\begin{equation}
{\cal G}(R,Z;a,z)=-2\sqrt{\frac{a}{R}} k \elik(k),
\label{eq:greenf}
\end{equation}
where
\begin{equation}
\elik(k) = \int_0^{\frac{\pi}{2}}{\frac{d\vartheta}{\sqrt{1-k^2 \sin^2\vartheta}}}
\label{eq:elik}
\end{equation}
is the complete elliptic integral of the first kind,
\begin{equation}
k=\frac{2\sqrt{aR}}{\Delta} \in [0,1]
\label{eq:k}
\end{equation}
is its modulus, and
\begin{equation}
\Delta^2=(R+a)^2+\zeta^2,
\end{equation}
where  $\zeta=Z-z$, and $(a,z)$ are the cylindrical coordinates of any point P' belonging to the shell. Basically, (\ref{eq:greenf}) corresponds to the potential created by an infinitesimally thin circular ring with unit mass per unit length, radius $a$ and altitude $z$. This function is known to be logarithmically singular at the location of the ring (where $k \rightarrow 1$). The gravitational potential generated, at any point P($\vec{r}$) of space, by such the axisymmetric shell is then given by the integral
\begin{flalign}
 \Psi(\vec{r}) = -2G \int_0^{2\pi} { \Sigma(\ell) \sqrt{\frac{a}{R}}  k \elik(k) d \ell},
\label{eq:psishell}
\end{flalign}
where $\Sigma$ is the local surface density, and $d \ell$ is the infinitesimal length along the shell section. In the case of a shell with a circular section of radius $\rc$, $a$ and $z$ are simply given by
\begin{flalign}
  a=\rc + b \cos \theta,
  \qquad
  z=b \sin \theta,
\label{eq:az}
\end{flalign}
where $\theta \in [0,2\pi]$ is the angular position of any point P' on the shell with respect to the equatorial plane (see Fig. \ref{fig:tshell.eps}). The infinitesimal length then takes its simplest form, namely $d \ell = b d\theta$. Other options are possible, but the subsequent calculations are much more complicated (see Sect. \ref{sec:comments}).

\begin{figure}
\includegraphics[width=6.1cm,bb=120 105 554 760,clip=true,angle=-90]{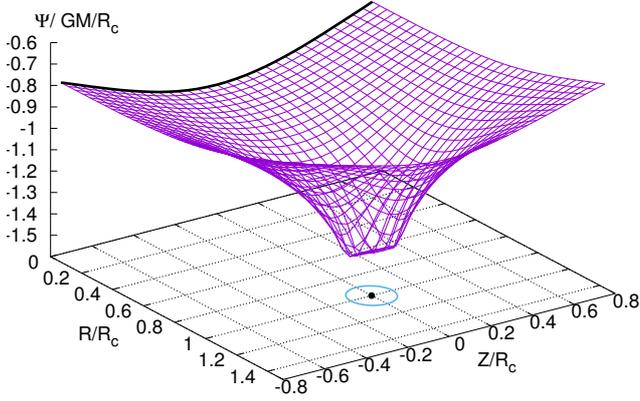}
\caption{The gravitational potential of the toroidal shell in units of $GM/\rc$ (homogeneous case) obtained by direct numerical computation of the integral in (\ref{eq:psishell}). The axis are in units of the main radius $\rc$. The normalized core radius (or shell parameter) is $e\equiv b/\rc=0.1$. Also shown are the values on the $Z$-axis from the formula by S{\'a}cha \& Semer{\'a}k (2005) ({\it thick black line}), the projected shell section ({\it blue line}) and its centre C ({\it black dot}); see also Fig. \ref{fig:tshell.eps}.}
\label{fig:pot.ps}
\end{figure}

The surface density $\Sigma$ may be variable in local latitude $\theta$. However, even if it is independent of $\theta$, (\ref{eq:psishell}) cannot in general be integrated into a compact form, except on the $Z$-axis \citep{sachas05}. An example of a direct numerical estimate of $\Psi$ is given in Fig. \ref{fig:pot.ps} for $e=0.1$. We use the trapezoidal rule as the quadrature scheme. We will use such a numerical potential, denoted $\psiref$ in the following, as a ``reference'' against which we will compare our approximations. As shown in \cite{htkl19}, the potential inside the shell cavity is a quasi-linear function of the cylindrical radius $R$, and it is weakly sensitive to the $Z$-coordinate, especially when $e \ll 1$. We see that the potential outside the cavity has a more complex structure. It resembles the potential of a loop, as already pointed out \citep[e.g.][]{wong73,ba11,kondratyev18}.

\section{Expansion of the Green function. Zero-order formula}
\label{sec:expansion}

As quoted in the introduction, the elliptic integral $\elik$ may be expanded over $k$ at $k \rightarrow 0$, which means far away from the source or close to the $Z$-axis, and over $k'=\sqrt{1-k^2}$ at $k \rightarrow 1$, e.g. close to or even inside the source \citep[see e.g.][]{as70,gradryz07}. However, such two series have to be matched somewhere \citep{ba11}. In the present paper, we propose a more synthetic approach which consists in expanding the axisymmetric Green function over $a$ and $z$, before integration over $\theta$ in (\ref{eq:psishell}). We expect to preserve the asymptotic behaviour of the potential at large distances. Let us remind that, for any ``regular enough'' function $f$ of two independent variables $x$ and $y$, the bivariate Taylor expansion at $(x_0,y_0)$ writes
\begin{flalign}
\label{eq:taylor2}
& f(x,y) = f(x_0,y_0)\\
&+ \sum_{n=1}^\infty\frac{1}{n!}\left\{\left[(x-x_0) \frac{\partial}{\partial x'} + (y-y_0) \frac{\partial}{\partial y'} \right]^n f(x',y') \right\}_{\begin{subarray}{l}x'=x_0\\y'=y_0\end{subarray}}.&
  \nonumber
\end{flalign}
The expansion is performed in $x\equiv a$ and $y\equiv z$, at the centre C of the shell, i.e. at $x_0\equiv\rc$ and $y_0\equiv 0$ (see Sect. \ref{sec:comments} for the expansion at the focal ring). We see from (\ref{eq:az}) that this is valid for $e < 1$, which in astrophysical toroids (typically orbiting a massive central body) is safely satisfied.

In fact, it is not necessary to expand the whole Green function. In particular, the term $\sqrt{a}$ is not problematic and it can be left aside. There are several options. For instance, if we expand $k \elik(k)$ or $\elik(k)$, the subsequent integration over $\theta$ will generate a new series of elliptic integrals (again, see Sect. \ref{sec:comments}). That is not a problem per se, but it complicates the calculations when the solid torus is considered. Seeing that the complication can be avoided when extracting the other factor $\sqrt{a}$ contained in the modulus (\ref{eq:k}), we finally choose to expand
\begin{flalign}
  \frac{\elik(k)}{\Delta} \equiv \newkappa
\end{flalign}
as
\begin{flalign}
\label{eq:taylor2kelik}
  & \newkappa =  \left. \newkappa\right|_{\begin{subarray}{l}a=\rc\\z=0\end{subarray}}+ (a-\rc) \left. \frac{\partial \newkappa }{\partial a} \right|_{\begin{subarray}{l}a=\rc\\z=0\end{subarray}}\\
  \nonumber
  & \quad + z \left. \frac{\partial \newkappa}{\partial z}\right|_{\begin{subarray}{l}a=\rc\\z=0\end{subarray}} + \frac{1}{2}(a-\rc)^2 \left. \frac{\partial^2  \newkappa}{\partial a^2} \right|_{\begin{subarray}{l}a=\rc\\z=0\end{subarray}}\\
  \nonumber
 & \quad\quad + (a-\rc)z \left. \frac{\partial^2 \newkappa}{\partial a \partial z} \right|_{\begin{subarray}{l}a=\rc\\z=0\end{subarray}} +  \frac{1}{2} z^2 \left. \frac{\partial^2 \newkappa}{\partial a \partial z} \right|_{\begin{subarray}{l}a=\rc\\z=0\end{subarray}}\\
  \nonumber
 & \quad\quad\quad+ \dots.
\end{flalign}

Note that $\newkappa$ is nothing but  the axisymmetric Green function, i.e. $\oint{|\vec{r}-\vec{r}'|^{-1}d \phi}$. Since it is a function of $a$, $z$, $R$ and $Z$, the infinite series is a polynomial (of ``infinite'' degree) in $a$ and $z$, whose coefficients are functions of $R$ and $Z$. This series naturally exhibits powers of the shell parameter $e$ which come from the partial derivatives and from the terms $a-\rc$ and $z$ as well; see (\ref{eq:az}). With (\ref{eq:taylor2kelik}), (\ref{eq:psishell}) becomes, at the lowest (zeroth) order,
\begin{flalign}
  \Psi(\vec{r}) \approx - 4G \Sigma_0 \newkappao b \times 2 \pi \rc S_{0,0} \equiv \Psi_0(\vec{r}),
\label{eq:psiorder0}
\end{flalign}
 where
\begin{equation}
\ko^2=\frac{4 \rc R}{\deltao^2},
\label{eq:ko}
\end{equation}
\begin{equation}
  \deltao^2=(R+\rc)^2+Z^2,
\end{equation}
\begin{equation}
  \newkappao = \frac{\elik(\ko)}{\Delta_0},
\end{equation}
and
\begin{equation}
  S_{0,0} = \frac{1}{2 \pi \Sigma_0 \rc}\int_{0}^{2 \pi}{\Sigma(\theta) a d\theta}
  \label{eq:s0}
\end{equation}
called the ``surface factor'' in the following. In this paper, we will consider {\em homogeneous} shells, so we set $\Sigma=$const.$=\Sigma_0$. Anticipating higher orders, let us define the whole series of definite integrals
\begin{flalign}
\label{eq:jnm}
  J_{n,m}& = \frac{1}{2 \pi}\int_0^{2\pi}{ \cos^n \theta \sin^m \theta  d\theta},
\end{flalign}
where $n$ and $m$ are positive integers. They can all be written in terms of the complete Beta function $B(\frac{n+1}{2},\frac{m+1}{2})$; see the Appendix \ref{sec:jnm}. We give $ J_{n,m}$ for the first few values of $n$ and $m$ in Tab. \ref{tab:jnm}. We note in particular that $J_{n,m}=0$ when either $m$, or $n$ or $m+n$ is odd. We have $S_{0,0} =  J_{0,0}+e J_{1,0} = 1$ and so the zero-order approximation for the potential of the shell is approximately given by
\begin{flalign}
 \Psi_0(\vec{r}) = - 8 \pi G\Sigma_0 b \rc \newkappao.
\label{eq:psiorder0final}
\end{flalign}

\begin{table}
  \centering
  \begin{tabular}{cccrcc}
 $n+m$ & $n$ & $m$ & $J_{n,m}$ & $S_{n,m}$ & $V_{n,m}$ \\ \hline
   $0$ & $0$ & $0$ & $\frac{1}{\pi} B(\frac{1}{2},\frac{1}{2})=1$ & $1$ & $1$\\\\
   $1$ & $1$ & $0$ & $0$ & $\frac{1}{2}e^2$ & $\frac{1}{4}e^2$\\
       & $0$ & $1$ & $0$ & \\\\
   $2$ & $2$ & $0$ & $\frac{1}{\pi} B(\frac{3}{2},\frac{1}{2})=\frac{1}{2}$ & $\frac{1}{2}e^2$ & $\frac{1}{4}e^2$\\
       & $1$ & $1$ & $0$ \\
       & $0$ & $2$ & $\frac{1}{\pi} B(\frac{1}{2},\frac{3}{2})=\frac{1}{2}$ & $\frac{1}{2}e^2$ & $\frac{1}{4}e^2$\\\\
   $3$ & $3$ & $0$ & $0$\\
       & $2$ & $1$ & $0$\\
       & $1$ & $2$ & $0$\\\hline
  \end{tabular}
  \caption{Expressions for $J_{n,m}$ required when expanding the Green function up to second order. Also given are the surface factor $S_{n,m}$ and the volume factor $V_{n,m}$.}
  \label{tab:jnm}
\end{table}

\subsection{Comparison with the potential of a point mass}
\label{subsec:similarity}

\begin{figure}
  \includegraphics[width=8.5cm,bb=0 0 477 411,clip=true]{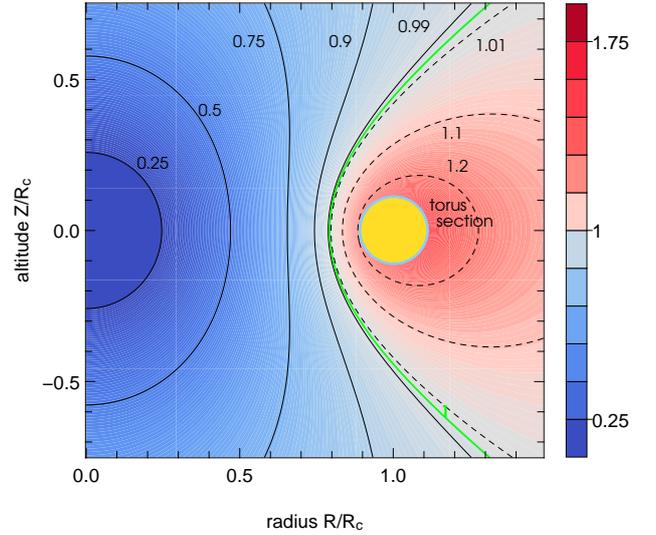}
  \caption{The factor $g_{0,0}$ given by (\ref{eq:f00}) and representing the deviation between the potential of a monopole and the zero-order potential of the toroidal shell. The conditions are the same as for Fig. \ref{fig:pot.ps}. The shell section is indicated ({\it thick black line}). A few contour lines are given : $g_{0,0}<1$ ({\it blue domain}), $g_{0,0}=1$ ({\it green line}), and $g_{0,0}>1$ ({\it red domain}).}
\label{fig:fclose.ps}
\end{figure}

It is easy to compare (\ref{eq:psiorder0final}) to the potential of some simple sources, like the potential of a monopole (or point mass at the origin), which is of major interest in dynamical studies. Introducing the mass of the homogeneous shell $\massshell=4 \pi^2 \Sigma_0 b \rc$, (\ref{eq:psiorder0final}) writes
\begin{flalign}
\Psi_0(\vec{r}) = - \frac{G \massshell}{r} \times g_{0,0},
\label{eq:psiorder0modmonopole}
\end{flalign}
where $r=\sqrt{R^2+Z^2}$ is the spherical radius, and
\begin{flalign}
g_{0,0} =  \frac{r}{\Delta_0} \frac{2}{\pi}\elik(\ko).
\label{eq:f00}
\end{flalign}
We see that (\ref{eq:psiorder0final}) differs from the monopole potential only by the quantity $g_{0,0}$ which is a function of the position in space only. In the physical space, it also depends on $\rc$, but not of $e$, i.e. $g_{0,0} \equiv g_{0,0}(\vec{r};\rc)$. The meridional-plane contours of $g_{0,0}$ are shown in Fig. \ref{fig:fclose.ps} in the neighborhood of the shell. The contours are closed, except the $g_{0,0}=1$ one. The region where $g_{0,0}>1$ surrounds the shell section, while $g_{0,0}<1$ concerns the central region near the $Z$-axis. This kind of map is helpful for dynamical studies since it indicates very well the families of bounded and unbounded trajectories of test particles moving with a constant angular momentum (on equatorial or inclined toroidal orbits and on purely meridional orbits).

\subsection{Comparison with the potential of a circular loop}

Let us also compare the zero-order shell potential (\ref{eq:psiorder0final}) and that of a circular loop of radius $\rc$ and mass $\massloop = 2 \pi \lambda \rc$, which writes
\begin{flalign}
  \Psi^{\rm loop}(\vec{r}) &= -2 G \lambda \sqrt{\frac{a}{R}} \ko \elik(\ko) \\
  &= - \frac{G\massloop}{r} \underbrace{\frac{r}{\Delta_0} \frac{2}{\pi}\elik(\ko)}_{g_{0,0}}.
  \nonumber
\end{flalign}
We see that, at the zeroth order, $\Psi_0 = \Psi^{\rm loop}$ at any point P in space, and for any value of $e$, provided $\massshell=\massloop$. According to \cite{kondratyev18} (see also below), there is no term in the expansion led by $e$ and, more generally, by odd powers of $e$ (this is not guaranteed as soon as $\Sigma$ varies with $\theta$). This implies that $\Psi^{\rm shell} =  \Psi^{\rm loop}+ {\cal O}(e^2).$ We can thus conclude that (similarity theorem 1):\\

{\it a homogeneous toroidal shell of main radius $\rc$ and circular section generates, at the first order in the $e$-parameter, the same exterior potential as a circular loop of same radius $\rc$ and same mass.\\
}

This result has a few important consequences. First, the approximation thus behaves correctly at infinity and on the $Z$-axis as well ($\lim_{k \rightarrow 0}\elik(\ko)=\frac{\pi}{2}$ and $\ko/\sqrt{R}$ is finite at the $Z$-axis). Second, the gravitational acceleration inherits these properties, i.e. the similarity theorem also applies to $\vec{g}=-\nabla \Psi$ (see below). Third, the formula (\ref{eq:psiorder0final}) does not generate any residual mass distribution in space. This is easily verifiable by calculating the Laplacian of $\newkappao$ (see the Appendix \ref{eq:residual} for the demonstration), i.e.
\begin{flalign}
  \nabla^2 \Psi_0 = 4\pi G \rho^{\rm res}=0.
\end{flalign}
This property is also intrinsic to the interior solutions reported in \cite{htkl19}.

\section{Numerical tests. Domain of validity}
\label{sec:efficiency}

\begin{figure}
\includegraphics[width=8.7cm,bb=76 265 552 695,clip=true]{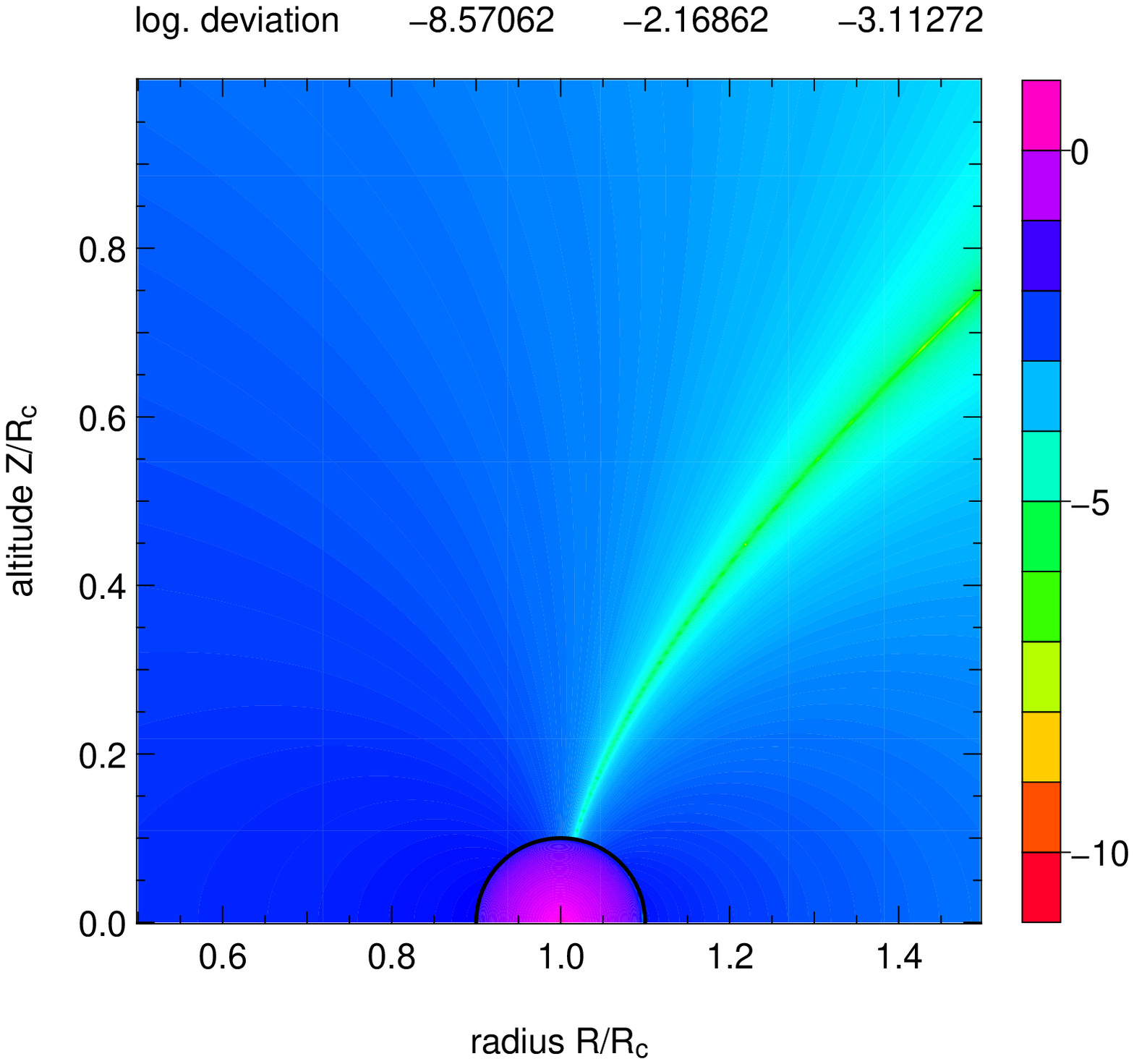}
\includegraphics[width=8.7cm,bb=76 265 552 695,clip=true]{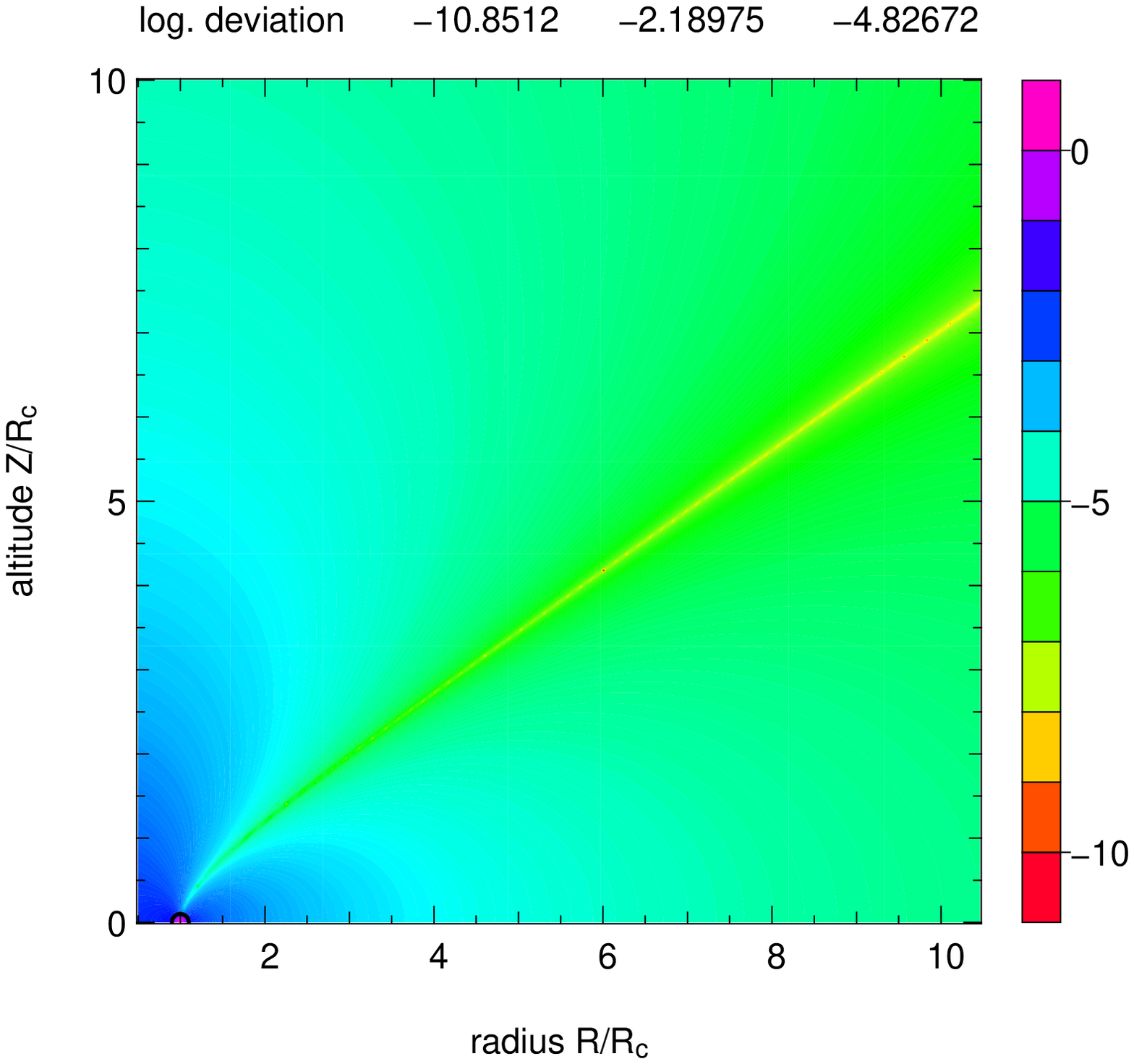}
\caption{The log. of the relative deviation defined by (\ref{eq:err_index}) between $\Psi$ computed by direct integration, i.e. $\psiref$, and the zero-order approximation given by (\ref{eq:psiorder0final}), in the vicinity of the shell ({\it top}) and at longer range ({\it bottom}). The parameter of the shell ({\it thick black circle}) is $e=0.1$; see Fig. \ref{fig:pot.ps} for the associated potential. The numbers given at the top, from left to right, refer to the minimal, maximal and mean values for $\epsilon$, respectively, reached within the actual computational box (and exterior to the shell).}
\label{fig:err0.ps}
\end{figure}

Let us now compare the expression (\ref{eq:psiorder0final}) to the numerical reference (see Sect. \ref{sec:potatpole}). We quantify the relative difference by
\begin{equation}
\epsilon = \log \left|\frac{\Psi - \psiref}{\psiref} \right|.
\label{eq:err_index}
\end{equation}
Figure \ref{fig:err0.ps} shows $\epsilon$ in the upper half-plane $Z>0$ for $e=0.1$. If we limit the statistics to the domain exterior to the shell, the average precision is of the order of $10^{-3}$ in the vicinity of the shell, and it is much lower in the far field. The deviation with respect to the reference never exceeds $1 \%$. The error is maximal near the surface of the shell, which is not a surprise. On the other hand, the best approximation is achieved in a narrow domain going from the top of the shell to infinity along the line $Z \sim 0.7 R$.

Our expansion is performed at the center C of the shell section. The expanded function thus has to be smooth enough between C and any point P’ located at the shell surface. This, however, is not the case for $\elik(k)/\Delta$ which is singular for any point P$(R,Z)$ belonging to the line segment [CP']. Therefore, the formula (\ref{eq:taylor2kelik}) and subsequently {\it the zero-order approximation is only valid outside the cavity}, namely for
\begin{equation}
(R-\rc)^2+Z^2 - b^2 >0.
\label{eq:outside}
\end{equation}

\begin{figure}
\includegraphics[height=8.5cm,bb=50 50 554 770,clip=true,angle=-90]{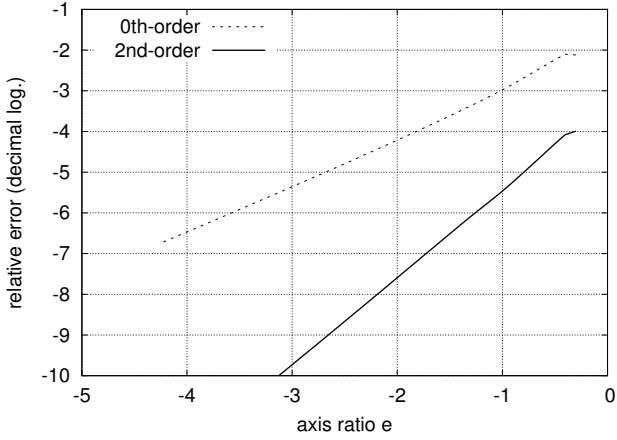}
\caption{Average of the log. of the relative deviation defined by (\ref{eq:err_index}) versus the shell parameter for the $0$th-order ({\it dotted line}) and the $2$nd-order approximation ({\it solid line}). The sample gather values contained inside the computational box $4e \times 4 e$ around the shell section (and exterior to it).}
\label{fig:err_vs012.ps}
\end{figure}

The comparison has been checked for different values of the shell parameter $e$. The results are plotted in Fig. \ref{fig:err_vs012.ps} where values of $\epsilon$ have been averaged over values contained inside a squared box $[1-2e,1+2e] \times [0,4e]$ (in dimensionless units) encompassing the shell section (see Fig. \ref{fig:err0.ps}); interior values are excluded. We see that the smaller the shell parameter, the better the approximation. The precision of the zero-order approximation remains better than $1\%$ for shell parameter as large as about $0.3$, which is remarkable.

\section{Expansion up to 2nd order}
\label{sec:orders12}

Though already very good, the zero-order approximation can be improved by considering further terms in the expansion. For order one, we have to calculate
\begin{flalign}
 \int_0^{2\pi}{\left[ (a-\rc) \left.\frac{\partial \newkappa}{\partial a}\right|_{\begin{subarray}{l}a=\rc\\z=0\end{subarray}}\right. }\left. +z \left.\frac{\partial \newkappa}{\partial z}\right|_{\begin{subarray}{l}a=\rc\\z=0\end{subarray}}\right] a b d\theta,
\end{flalign}
where $a$ and $z$ are still given by (\ref{eq:az}). Because the derivatives are evaluated at $a = \rc$ and $z=0$, they are not concerned by the integration over $\theta$ and can be carried out of the operator. There are two new surface factors to calculate, namely (we do not include $\Sigma$ in these definitions, since we assume it is constant)
\begin{flalign}
  S_{1,0} =  \frac{1}{2\pi\rc^2}\int_0^{2\pi}{(a-\rc)a d\theta},
\end{flalign}
and
\begin{flalign}
  S_{0,1} =  \frac{1}{2\pi\rc^2}\int_0^{2\pi}{za d\theta},
\end{flalign}
but this latter term vanishes (since $J_{0,1}$ and $J_{1,1}$ are zero). In the first order, the potential writes $\Psi \approx \Psi_0+ \Psi_1$, where $\Psi_0$ is given by (\ref{eq:psiorder0final}) and 
\begin{flalign}
\Psi_1 = -8 \pi G \Sigma_0 b \rc^2 \left.\frac{\partial \newkappa}{\partial a}\right|_{\rc,0} \times S_{1,0},
\label{eq:firstorder}
\end{flalign}
where $S_{1,0}=e(J_{1,0} + e J_{2,0})$. Note that the derivative $\frac{\partial \newkappa}{\partial a}$ is analytical (see the Appendix \ref{sec:derivatives_kk}). Since $J_{1,0}=0$, the first-order correction depends on $e^2$. As $2 \pi J_{2,0}=2B(\frac{3}{2},\frac{1}{2})=\pi$, we have
\begin{flalign}
  S_{1,0} = \frac{1}{2}e^2.
\end{flalign}

It is pertinent to account for the next term in the Taylor expansion which also contains a contribution varying as $e^2$. This term is
\begin{flalign}
  \nonumber
  \frac{1}{2}&\int_0^{2\pi}{\left[  (a-\rc)^2 \left.\frac{\partial^2 \newkappa}{\partial a^2}\right|_{\begin{subarray}{l}\rc\\0\end{subarray}}+2 (a-\rc)z\left.\frac{\partial^2 \newkappa}{\partial a \partial z}\right|_{\begin{subarray}{l}\rc\\0\end{subarray}} \right.}\\
      & \qquad\qquad +  \left.  z^2 \left.\frac{\partial^2 \newkappa}{\partial z^2}\right|_{\begin{subarray}{l}\rc\\0\end{subarray}}\right] a b d\theta,
\end{flalign}
and so the $2$nd-order approximation is given by $\Psi \approx \Psi_0+ \Psi_1+ \Psi_2$, with
\begin{flalign}
\label{eq:secondorder} 
& \Psi_2 = - 8  \pi G\Sigma_0 b \rc^3 \\
  \nonumber
  &\times  \frac{1}{2} \left[ \left.\frac{\partial^2 \newkappa}{\partial a^2} \right|_{\begin{subarray}{l}\rc\\0\end{subarray}} S_{2,0} +
    2 \left.\frac{\partial^2 \newkappa}{\partial a \partial z}\right|_{\begin{subarray}{l}\rc\\0\end{subarray}}S_{1,1} 
    +  \left.\frac{\partial^2 \newkappa}{\partial z^2}\right|_{\begin{subarray}{l}\rc\\0\end{subarray}} S_{0,2} \right],
  \nonumber
\end{flalign}
where the derivatives are given in the Appendix \ref{sec:derivatives_kk}. The new surface factors are (again, $\Sigma$ is removed from these definitions)
\begin{flalign}
  S_{2,0} =  \frac{1}{2\pi\rc^3}\int_0^{2\pi}{(a-\rc)^2a d\theta},
\end{flalign}
\begin{flalign}
  S_{1,1} =  \frac{1}{2\pi\rc^3}\int_0^{2\pi}{(a-\rc)za d\theta},
\end{flalign}
and
\begin{flalign}
  S_{0,2} =  \frac{1}{2\pi\rc^3}\int_0^{2\pi}{z^2a d\theta}.
\end{flalign}
We see that $S_{1,1}=0$, again because of the odd power of $z$. The non-zero terms are $S_{2,0} = e^2 (J_{2,0}+e  J_{3,0})$ and $S_{0,2} = e^2 (J_{0,2}+e J_{1,2})$, where $J_{3,0}= J_{1,2}=0$, $2 \pi J_{2,0}=2B(\frac{3}{2},\frac{1}{2})=\pi$. $S_{2,0}= S_{0,2} = \frac{1}{2}e^2$. We thus see that it is necessary to include {\em both} orders $1$ and $2$ simultaneously in order to obtain a consistent $e^2$-approximation. This new approximation can be put in the form of a modified monopole, like we did for the zeroth-order expression. There is one specific correction factor $g_{n,m} \equiv g_{n,m}(\vec{r};\rc,e)$ for each non-zero surface factor $S_{n,m}$. We find
\begin{flalign}
  \Psi_1 = - \frac{G \massshell}{r} g_{1,0},
\end{flalign}
where
\begin{flalign}
  g_{1,0}=r\frac{2}{\pi} \left. \frac{\partial \newkappa}{\partial {a}} \right|_{\begin{subarray}{l}a=\rc\\z=0\end{subarray}} \rc S_{1,0},
  \end{flalign}
and
\begin{flalign}
  \Psi_2 = - \frac{G\massshell}{r} \frac{1}{2} \left( g_{2,0} + g_{0,2}\right),
\end{flalign}
where
\begin{flalign}
  g_{2,0}=r\frac{2}{\pi} \left. \frac{\partial^2 \newkappa}{\partial {a}^2} \right|_{\begin{subarray}{l}a=\rc\\z=0\end{subarray}} \rc^2 S_{2,0},
  \end{flalign}
and
\begin{flalign}
  g_{0,2}=r \frac{2}{\pi} \left. \frac{\partial^2 \newkappa}{\partial {z}^2} \right|_{\begin{subarray}{l}a=\rc\\z=0\end{subarray}} \rc^2 S_{0,2}.
\end{flalign}

Because of the circular section, $S_{2,0}=S_{0,2}$ which implies that the partial sum $\Psi_1+\Psi_2$ can be rewritten in a very compact form. We actually find
  \begin{flalign}
    \nonumber
    g_{1,0} +& \frac{1}{2} \left( g_{2,0} + g_{0,2}\right) =  r \frac{2}{\pi} \times \frac{e^2}{8\kp^2 \Delta_0^3}\\
    &\times \left\{ \left[\Delta_0^2-2\rc(\rc+R)\right] \elie(k) - \kp^2 \Delta_0^2 \elik(k)\right\},
  \end{flalign}
 which is to be multiplyed by $-G\massshell/r$.
\begin{figure}
\includegraphics[height=7.7cm,bb=76 265 556 695,clip=true]{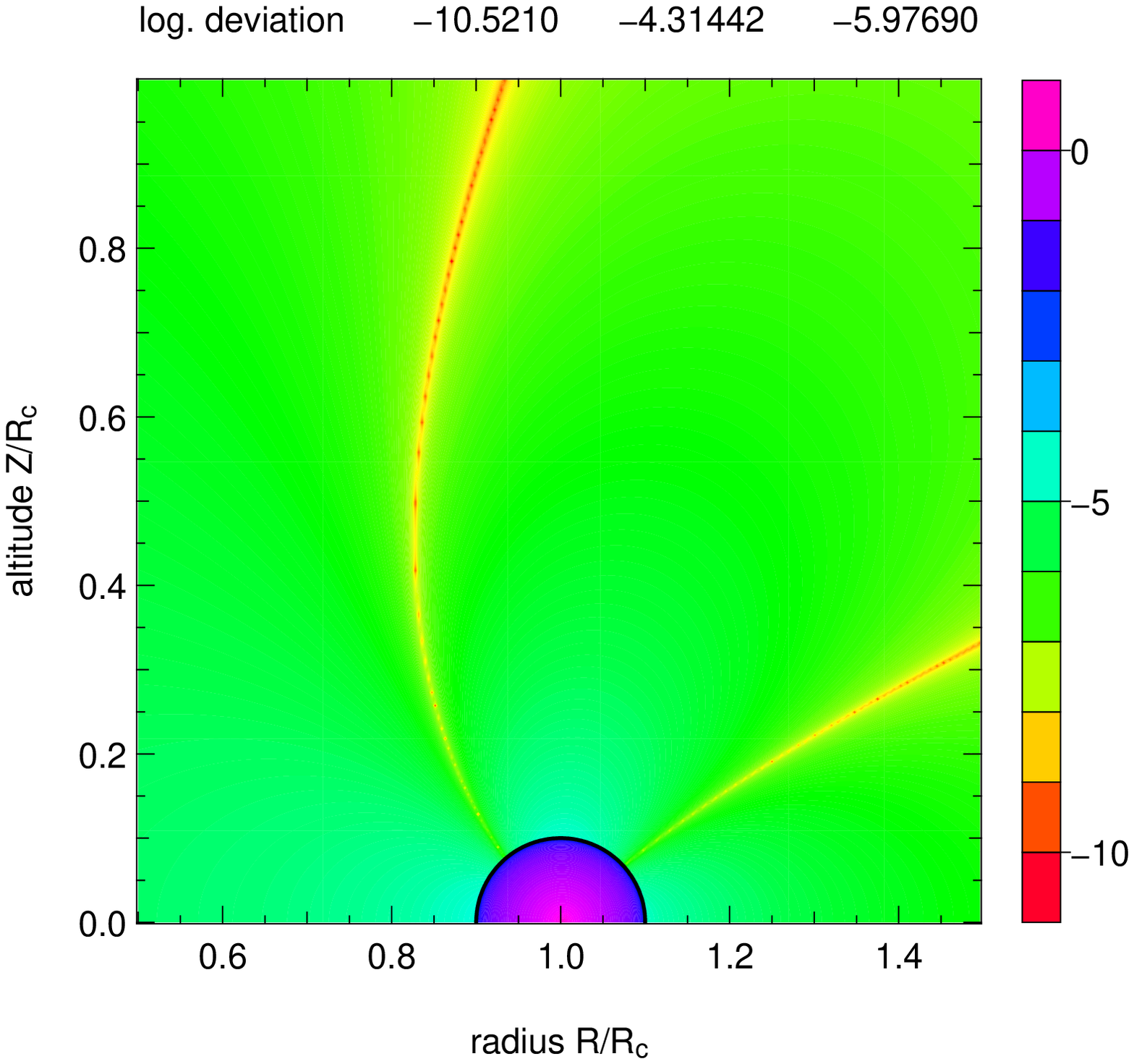}
\includegraphics[height=7.7cm,bb=76 265 556 695,clip=true]{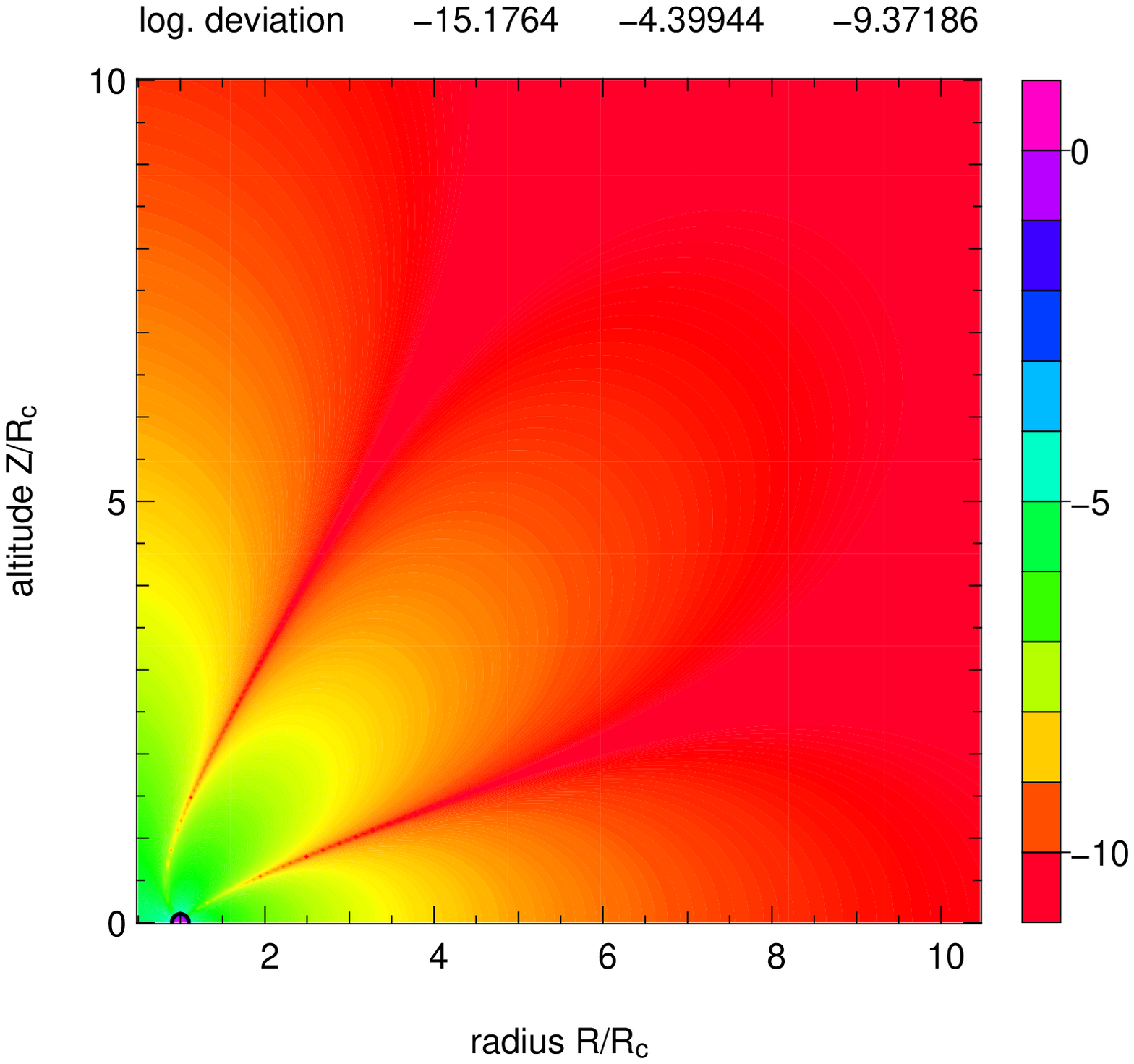}
\caption{The same legend and the same conditions as for Fig. \ref{fig:err0.ps}, but for the $2$nd-order approximation (i.e. $e^2$-approximation).}
\label{fig:psi_err2.ps}
\end{figure}

Figure \ref{fig:psi_err2.ps} compares the second-order approximation obtained from (\ref{eq:psiorder0final}), (\ref{eq:firstorder}) and (\ref{eq:secondorder}) with the reference values (see Sect. \ref{sec:potatpole}), as computed under the same conditions as in Fig. \ref{fig:err0.ps}. We notice that the $e^2$-approximation reproduces the potential with almost $6$-digits precision in the close vicinity of the shell. At larger distances, the expansion is extremely efficient (in the present example, the potential is known with more than $10$ digits for $r/\rc \gtrsim 5$ typically). The variation of the averaged precision as a function of the shell parameter $e$ is plotted in Fig. \ref{fig:err_vs012.ps}, in the same conditions as for the zeroth-order approximation.

\section{Values at the surface. Match with the interior solution}
\label{eq:connection}

\begin{figure}
\includegraphics[height=8.5cm,bb=50 50 554 770,clip=true,angle=-90]{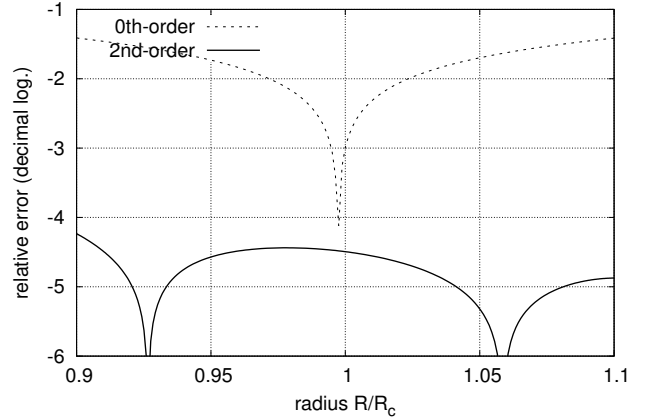}
\caption{Logarithm of the difference between the interior solution \citep{htkl19} and the exterior solution for the $0$th-order ({\it dotted line}) and $2$nd-order approximations ({\it plain line}). The shell parameter is $e=0.1$.}
\label{fig:err_gamma.ps}
\end{figure}

At the surface of the shell, we have $R=\rc+b \cos \theta$ and $Z=b \sin \theta$, or equivalently $(R-\rc)^2+Z^2- b^2=0$ with $R \in [\rc-b,\rc+b]$. If we introduce these expressions in (\ref{eq:ko}), the modulus $\ko$ simplifies into 
\begin{flalign}
  \ko^2 = \frac{4R\rc}{4R\rc+b^2}.
\end{flalign}
Using these values in (\ref{eq:psiorder0final}), (\ref{eq:firstorder}) and (\ref{eq:secondorder}), we get the potential at the shell surface, which can be compared to the values obtained from the interior solution reported in \cite{htkl19}; see their equations (35), (36) and (39). The results are displayed in Fig. \ref{fig:err_gamma.ps}, again for $e=0.1$. We see that the matching is very good at the second order (with at least $4$ correct digits). At the zeroth order, the interior solution reduces to a constant potential throughout the toroidal cavity, which is quite crude while the exterior solution already depends on the radius $R$.

\section{Generalization to $n$th-order expansion}
\label{sec:gen}

It is possible to include further terms in the Taylor series. Since the expansion writes formally
\begin{flalign}
\label{eq:taylor2kelikexpnded}
  \newkappa&= \newkappao&\\
  \nonumber
  &+ \sum_{n=1}^\infty \frac{1}{n!}  {\sum_{m=0}^n {\binom{n}{m}(a-\rc)^{n-m}z^m  \left. \frac{\partial^n \newkappa}{\partial {a}^{n-m} \partial {z}^m} \right|_{\begin{subarray}{l}a=\rc\\z=0\end{subarray}} }},
  \nonumber
\end{flalign}
where $\binom{n}{m}$ denotes the binomial coefficient, the potential can be exactly reconstructed by multiplying (\ref{eq:taylor2kelikexpnded}) by $a$, followed by the integration over the latitude angle $\theta$. If the infinite series is truncated at order $N$, the potential is of the form
\begin{flalign}
  \Psi \approx \Psi_0 + \sum_{n=1}^N{\Psi_n},
  \label{eq:psiNseries}
\end{flalign}
where the $n$th-order contribution $\Psi_n$ is made of $n+1$ terms, namely
\begin{flalign}
\label{eq:dpsin}
\Psi_n &= -4G\Sigma_0 \\
&\times \frac{1}{n!} \sum_{m=0}^n \binom{n}{m}  \left. \frac{\partial^n \newkappa}{\partial {a}^{n-m} \partial {z}^m} \right|_{\begin{subarray}{l}a=\rc\\z=0\end{subarray}} \times 2 \pi b \rc^{n+1} S_{n-m,m},
  \nonumber
\end{flalign}
where we have set (still in the homogeneous case)
\begin{flalign}
  \label{eq:snm}
  S_{n,m} & =  \frac{1}{2\pi \rc^{n+m+1}}\int_0^{2\pi}{(a-\rc)^n z^m a d\theta}\\
  \nonumber
  &=e^{n+m}(J_{n,m}+eJ_{n+1,m}).
\end{flalign}
Note that $\Psi_{n+1} \ll \Psi_n$ when $e \ll 1$, and the equality in (\ref{eq:psiNseries}) is obtained in the limit $N \rightarrow \infty$. In the form of the modified monopole representation, the $n$-order correction is 
\begin{flalign}
\Psi_n = - \frac{G \massshell}{r} \sum_{m=0}^n{g_{n,m}},
\label{eq:dspsintorus}
\end{flalign}
where
\begin{flalign}
g_{n,m} =  \frac{2}{\pi} r \times \frac{1}{n!} & \binom{n}{m}  \left. \frac{\partial^n \newkappa}{\partial {a}^{n-m} \partial {z}^m} \right|_{\begin{subarray}{l}a=\rc\\z=0\end{subarray}} \rc^n S_{n-m,m}.
\end{flalign}

Since the two operators $\nabla^2_{R,Z}$ and $\partial^n / \partial {a}^{n-m} \partial {z}^m$ act on different spaces, we have
\begin{flalign}
  \nonumber
  \nabla_{R,Z}^2 \left(\frac{\partial^n \newkappa}{\partial {a}^{n-m} \partial {z}^m} \right)& = \frac{\partial^n}{\partial {a}^{n-m} \partial {z}^m} \left( \nabla_{R,Z}^2 \newkappa \right)\\
  & =0
\end{flalign}
for any pair $(n,m)$. This is expected because $\newkappa$ is a harmonic function (see Sec. \ref{sec:efficiency}). We thus conclude that $\nabla_{R,Z}^2 \Psi_n=0$, for any $n$, which means that each term of the expansion separately obeys the Laplace equation. Therefore, expanding the Green function over $a$ and $z$ induces no residual source mass in space, whatever the order of the truncation.

\begin{table}
  \centering
  \begin{tabular}{cccccc}
      &     &     & $V_{n,m}$ \\ 
$n+m$ & $n$ & $m$  & homogeneous & (\ref{eq:rhobprim}) \\ \hline
   $0$ & $0$ & $0$ & $1$ & $\frac{\alpha}{\alpha+1}$\\\\
   $1$ & $1$ & $0$ & $\frac{1}{4}e^2$ & $\frac{\alpha}{4(\alpha+2)}e^2$\\
       & $0$ & $1$ \\\\
   $2$ & $2$ & $0$ & $\frac{1}{4}e^2$ & $\frac{\alpha}{4(\alpha+2)}e^2$\\
       & $1$ & $1$ & $0$ & $0$ \\
       & $0$ & $2$ & $\frac{1}{4}e^2$ & $\frac{\alpha}{4(\alpha+2)}e^2$\\\hline
  \end{tabular}
  \caption{Expressions for the volume factor $V_{n,m}$ in the case of core-stratified toroids according to (\ref{eq:vnminhomo}) and (\ref{eq:rhobprim}).}
  \label{tab:jnminhomo}
\end{table}

\section{The solid torus}
\label{sec:solidtorus}

The above solution for the shell can be employed to obtain the potential of a solid toroid. This is achieved by integrating (\ref{eq:psishell}) over $b$, while the surface density $\Sigma$ is changed for $\rho db$, $\rho$ being the mass density. The result is
\begin{flalign}
 \Psi(\vec{r}) = -4G \int_0^b{\int_0^{2\pi} { \rho(b',\theta) a \kappa b' d\theta} db'},
\label{eq:psisolidtorus}
\end{flalign}
where $\newkappa$ can be replaced by its Taylor expansion, namely (\ref{eq:taylor2kelik}). In the leading term, i.e. using just (\ref{eq:psiorder0final}), we have
\begin{flalign}
  \Psi_0(\vec{r}) = & - 4 G \newkappao \times \pi \rc b^2 \rho_0 V_{0,0},
\label{eq:psisolidtorusorder0}
\end{flalign}
where $\rho_0$ is some typical mass density, and $V_{0,0}$ is the ``volume factor'' defined in general by
\begin{flalign}
  \label{eq:volumefactor00}
  V_{0,0} & = \frac{1}{\pi \rho_0 \rc b^2}\int_{0}^{b}{ b' db'\int_{0}^{2 \pi}{\rho(b',\theta)a d \theta}},
\end{flalign}
where $b'=e' \rc \le b$. As quoted, the mass density $\rho$ may vary with both $\theta$ and $b'$ (see below). In the {\em homogenous case}, we have $\rho=$const.$=\rho_0$, and so (\ref{eq:volumefactor00}) becomes
\begin{flalign}
  \nonumber
  V_{0,0} & = \frac{2}{\rho_0 e^2} \int_{0}^{e}{\rho(e') e' S_{0,0}(e') de'}\\
  & = \frac{2}{e^2} \int_{0}^{e}{e' S_{0,0}(e') de'},
\end{flalign}
where the dependence of the surface factor with the shell parameter $e$ has been explicited. Introducing the total mass of the homogeneous torus $\masssolid= 2 \pi^2 \rho_0 b^2 \rc$, the zero-order formula can be written in the form
\begin{flalign} 
\Psi_0(\vec{r}) = - \frac{G\masssolid}{r} g_{0,0},
\label{eq:psisolidtorusorder0monopole}
\end{flalign}
where we have set
\begin{flalign}
g_{0,0} = \frac{r}{\deltao} \frac{2}{\pi}\elik(\ko) V_{0,0}.
\end{flalign}
Again, the difference from the point-mass potential is represented by the term $g_{0,0}$, while the deviation with respect to the potential of a massive loop (of radius $a=\rc$) is given by the volume factor $V_{0,0}$. For the zero-order approximation, we have $V_{0,0}=1$, and so $\Psi^{\rm solid} =  \Psi^{\rm loop}+ {\cal O}(e^2).$ We can thus conclude (similarity theorem 2):\\

\noindent {\it a solid torus of main radius $\rc$ and circular section generates, at the first order in the $e$-parameter, the same exterior potential as a circular loop of radius $\rc$ and same mass.}\\

The derivation of the $e^2$-term requires $V_{1,0}$, $V_{2,0}$ and $V_{0,2}$, which are listed in Tab. \ref{tab:jnminhomo}. These quantities happen to be equal (due to the circular section). As a consequence, the partial sum $\Psi_1+\Psi_2$ resembles\footnote{We notice two differences between (\ref{eq:secondorderkondratyev}) and the formula (14) by \cite{kondratyev18}: the factor $R_0^3$ should be $R_0 r_0^2 \equiv \rc b^2$ (since the $\phi_2$ term is multiplied by $e^2$), and the factor $16$ at the denominator should be $4$.} the formula derived in \cite{kondratyev18}. We finally find
   \begin{flalign}
   \label{eq:secondorderkondratyev}
     \Psi_1+\Psi_2  &=  e^2 \left(- \frac{G \pi \rho_0 \rc b^2}{4\kp^2 \Delta_0^3} \right.\\
    \nonumber
    &\left. \times \left\{ \left[\Delta_0^2-2\rc(\rc+R)\right] \elie(k) - \kp^2 \Delta_0^2 \elik(k)\right\} \right).
   \end{flalign}
   
We have compared (\ref{eq:psisolidtorusorder0}) to a reference obtained by direct numerical integration of (\ref{eq:psisolidtorus}). As we have observed, the error map is the same as for the shell, which is expected since the only difference between the shell and the torus stands in the volume factor which is analytical. This remark holds for the $e^2$-approximation.

More terms in the expansion of $\newkappa$ can be accounted for. The $n$-order contribution is
\begin{flalign}
\label{eq:dpsin}
\Psi_n &= -4G \frac{1}{n!} \sum_{m=0}^n \binom{n}{m}  \left. \frac{\partial^n \newkappa}{\partial {a}^{n-m} \partial {z}^m} \right|_{\begin{subarray}{l}a=\rc\\z=0\end{subarray}} \\
&\times 2 \pi \rc^{n+1} \rho_0 \int_0^b{S_{n-m,m}(b')b' db'},
  \nonumber
\end{flalign}
where $S_{n-m,m}$ depends on $b'$ as indicated. If we set the volume factor $V_{n,m}$ to
\begin{flalign}
  \nonumber
  V_{n,m} &=  \frac{1}{2\pi\rc^{n+m+1}} \frac{2}{b^2}\int_0^b{b' db'\int_0^{2\pi}{(a-\rc)^n z^m a d\theta}}.\\
  \label{eq:vnm}
  & = \frac{2}{e^2} \int_0^e{S_{n,m}(e') e' de'},
\end{flalign}
then $\Psi_n$ has the same form as (\ref{eq:dspsintorus}) where $S_{n-m,m}$ is just to be replaced by $V_{n-m,m}$ and $\massshell$ by $\masssolid$. It can be checked that $\Psi_n$ is harmonic.

\section{Inhomogenous systems}
\label{sec:inhomogeneous}

As a matter of fact, (\ref{eq:taylor2kelikexpnded}) works for inhomogenous systems. Actually, the expansion depends on $a$ and $z$ only through powers of $\cos \theta$ and $\sin \theta$. If $\Sigma(\theta)$ is prescribed, the knowledge of any term $\Psi_n$ just requires the calculation of the surface factors according to
\begin{flalign}
  S_{n,m} & =  \frac{1}{2\pi \Sigma_0 \rc^{n+m+1}}\int_0^{2\pi}{\Sigma(\theta)(a-\rc)^n z^m a d\theta}\\
  \nonumber
  & =  \frac{e^{m+n}}{2\pi \Sigma_0}\int_0^{2\pi}{\Sigma(\theta)\cos^n \theta \sin^m \theta(1 + e \cos \theta) d\theta}.
\end{flalign}
We see that the $S_{n,m}$'s are combinations of moments of the surface density profile, which are analytical for a wide family of $\Sigma(\theta)$-profiles. In a similary way for the  solid torus, if $\rho$ depends both on $\theta$ and on $b' \le b$, then the volume factors are calculated following
\begin{flalign}
  \nonumber
  &V_{n,m}  =  \frac{1}{2\pi \rho_0 \rc^{n+m+1}} \frac{2}{b^2}\\
\label{eq:vnminhomo}
  & \qquad \times \int_0^b{b' db'\int_0^{2\pi}{\rho(b',\theta)(a-\rc)^n z^m a d\theta}},\\
  \nonumber
  & \qquad =  \frac{e^{m+n}}{\pi \rho_0}\\
  \nonumber
  & \times \int_0^1{{x'}^{n+m+1} dx'\int_0^{2\pi}{\rho(x',\theta)\cos^n \theta \sin^m \theta(1 + b x' \cos \theta) d\theta}},
\end{flalign}
where we have set $b'=bx' \le b$. We can go a little bit further in the analysis by considering the case where the two variables $b$' and $\theta$ are separable, i.e. $\rho(b',\theta) = f(b') \times g(\theta)$. This corresponds to toroids having a core stratification. For instance, if we assume the $\theta$-invariance and (with $2\alpha > -1$)
\begin{flalign}
\rho(b')=\rho_0\left[1-\left(\frac{b'}{b}\right)^{2\alpha}\right],
\label{eq:rhobprim}
\end{flalign}
then the volume factor required at order zero (i.e. $n=m=0$) is
\begin{flalign}
  \nonumber
  V_{0,0} & = \int_0^1{2x'(1-{x'}^{2\alpha}) dx'}\\
  &= \frac{\alpha}{1+\alpha}.
\end{flalign}

Note that $V_{0,0} \rightarrow 1$ as $\alpha \rightarrow \infty$. Since $\masssolid V_{0,0}$ is just the total mass $M$ of the (inhomogeneous) core-stratified torus, we have $\Psi =  \Psi^{\rm loop}+ {\cal O}(e^2)$. We can conclude that (similarity theorem 3):\\

  \noindent {\it a core-stratified torus of main radius $\rc$ and circular section generates, at the first order in the $e$-parameter, the same exterior potential as a circular loop of radius $\rc$ and same mass.}\\

Table \ref{tab:jnminhomo} lists values of $V_{n,m}$ corresponding to (\ref{eq:rhobprim}).

\section{Gravitational acceleration}
\label{sec:acceleration}

\begin{figure}
  \includegraphics[width=8.6cm,bb=50 50 410 302,clip=true]{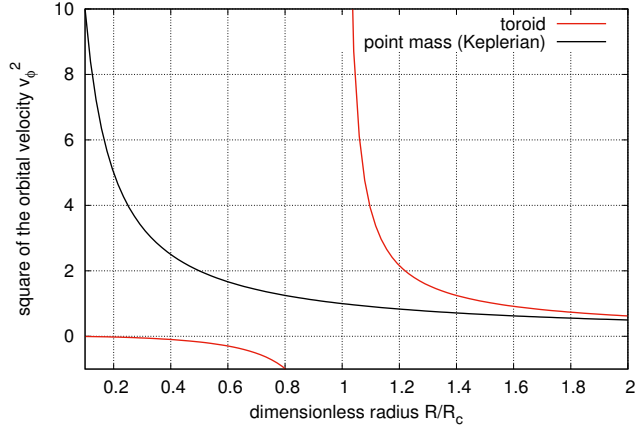}
\caption{Square of the circular velocity in the equatorial plane of the toroid as given by (\ref{eq:v2}), i.e. at order zero. The curve has to be truncated at the actual outer radius, which is $1+e$ in dimensionless units. Negative values take sense only when a massive central object is present. The Keplerian velocity due to a point mass with the same mass is shown in comparison.}
\label{fig:v2.ps}
\end{figure}

For a massive loop, the non-zero components of acceleration $\vec{g} = - \vec{\nabla} \Psi$ are  given by \citep{durand64,hure05}
\begin{flalign}
   \label{eq:gr}
  g_R &=  \frac{G \massloop}{2\pi \rc R} \sqrt{\frac{\rc}{R}} \\
  \nonumber
  &\qquad \quad \times \ko \left[ \elie(\ko)- \elik(\ko) +  \frac{(\rc-R)\ko^2 \elie(\ko)}{2 \rc \kpo^2}  \right],
\end{flalign}
and
\begin{flalign}
    g_Z = - \frac{ G\massloop Z}{4 \pi R \rc \sqrt{R\rc}} \frac{\ko^3 \elie(\ko)}{\kpo^2}.
   \label{eq:gz}
\end{flalign}
According to the similarity theorems 1 to 3, the acceleration outside a shell, solid torus, or core-stratified toroid is the same as for a loop having the same mass $M$, and deviations are ${\cal O}(e^2)$. This result is very convenient for a study of the motion of orbiting test-particles. Several types of trajectories can be distinguished. Of particular interest are circular trajectories tied to the equatorial plane of the shell/torus and having $R \notin [\rc-b,\rc+b]$. The orbital velocity $v_\phi^2 = R \nabla_R \Psi$ is easily deduced from (\ref{eq:gr}). We find
\begin{flalign}
  \label{eq:v2}
  v_\phi^2(R) =  \frac{GM}{R+\rc} \frac{1}{\pi} \left[ \elik(\ko) + \frac{R+\rc}{R-\rc} \elie(\ko) \right],
\end{flalign}
where $\ko=\frac{2\sqrt{R\rc}}{R+\rc}$ follows from (\ref{eq:ko}) where $Z$ has been set to $0$. Figure \ref{fig:v2.ps} displays (\ref{eq:v2}) versus the radius. Note that $v_\phi^2(R) \le 0 $ for $R \in [0,\rc-b]$, which means that orbits are in principle forbidden in this region, unless a massive central object is present. The Keplerian profile associated to a point mass at the origin is shown in comparison. For $R \ge \rc+b$, the velocity is super-Keplerian. It is a decreasing function of the radius, the maximum value being reached at the outer radius $\rc+b$ of the toroid. To the detriment of precision, we can replace the elliptic integrals by more standard functions when $\ko \rightarrow 1$, which corresponds to particles orbiting very close to the inner/outer radius of the toroid. Within this limit, $\elik(\ko) \sim \ln \frac{4(R+\rc)}{|R-\rc|}$ and $\elie(\ko) \sim 1$.

\begin{figure}
  \includegraphics[width=8.6cm,bb=50 50 410 302,clip=true]{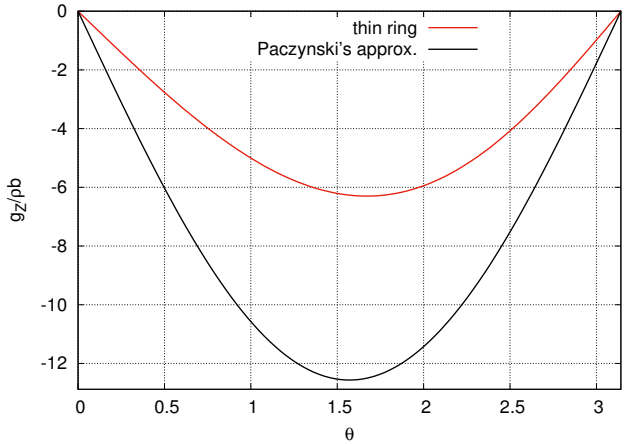}
\caption{Vertical acceleration at the surface of the toroid in units of $G\rho b$. Paczynski's approximation valid for geometrically thin, extended systems is shown in comparison.}
\label{fig:paczynski.ps}
\end{figure}

Another interesting quantity is the vertical component of acceleration at the surface of thin/small rings. It is a fundamental ingredient that governs the hydrostatic equilibrium of astrophysical discs \cite[e.g.][]{ss73,pringle81}. By setting $\elie(\ko) \approx 1$, which again corresponds to the vicinity of the toroid, we find
\begin{flalign}
  g_Z = - 2 \pi G \rho b \frac{\sin \theta}{\sqrt{1+e \cos \theta + \frac{e^2}{4}}},
  \label{eq:gz-nopac}
\end{flalign}
where $\theta \in [0,\pi]$ above the equatorial plane. This quantity is plotted in Fig. \ref{fig:paczynski.ps} for $e=0.1$ as the torus parameter. It varies between $0$ at the inner/outer edges to about $-2 \pi G \rho b$ at $\theta=\frac{\pi}{2}$. It is interesting to see that Paczynski's approximation \citep{pacz78}, classically written as $-4 \pi G \rho Z$, overestimates the acceleration by a factor $2$ in the middle of the toroid. This observation may e.g. be of importance in oscillation modes in planetary or other rings \citep{wt88,lss19}. It also means that, in a geometrically thin discs where Paczynski's approximation is valid, half of the vertical acceleration comes from the local contribution of matter while the other half comes from the global or long-range distribution of matter \citep{thh2014aa}. 

\section{Magnetic potential and field for purely azimuthal currents (in surface and volume)}
\label{sec:ab}

The method presented in this paper can also be applied to the determination of the vector potential $\vec{A}$ of electromagnetism. Toroidal currents are met in both terrestrial and astrophysical plasmas \citep{dbak09,audrey18}. The magnetic potential $\vec{A}=A_\phi\vec{e}_\phi$ of a toroidal shell carrying a purely azimuthal electric current $\sigma \vec{e}_\phi$ is obtained by summing over the contribution of individual current loops \citep{jackson98,cohl01}, namely
\begin{equation}
A_\phi(\vec{r}) = \frac{\mu_0}{2\pi} \int_0^{2\pi}{ \sigma(\theta)\sqrt{\frac{a}{R}}\frac{\left(2-k^2\right) \elik(k) -2 \elie(k)}{k} b d\theta} \,,
\end{equation}
where $I= b \oint{\sigma(\theta) d\theta}$ is the total current. Similarly as for the gravitational problem, we have to select some part of the Green function. A convenient choice appears to be
\begin{flalign}
 \frac{1}{\Delta}\left\{\frac{2}{k^2}\left[\elik(k) - \elie(k)\right] -\elik(k)\right\} \equiv \newkappa'.
\end{flalign}
By expanding $\newkappa'$ over $a$ and $z$ at the center C of the shell, i.e. at $a=\rc$ and $z=0$, and integrating over the latitude $\theta$ (see Sec. \ref{sec:expansion}), we get the leading term
\begin{flalign}
  \nonumber
  A_\phi(\vec{r}) & = \frac{\mu_0  b  }{\pi} \times \newkappa'_0 \int_{0}^{2 \pi}{\sigma a d\theta}\\
  &= 2 \mu_0 \sigma b \rc \newkappa'_0 S_{0,0},
\label{eq:aorder0}
\end{flalign}
where $\newkappa'_0$ stands for $\newkappa'$ evaluated at C, $\sigma=$const. is assumed, and $S_{0,0}$ is given by (\ref{eq:s0}). A $\theta$-dependent surface density of current would lead to a different surface factor. Again, we notice that (\ref{eq:aorder0}) formally differs from the expression for a current loop only by the term $S_{0,0}$, which is unity in the homogeneous case. We thus state (similarity theorem 4):\\

\noindent {\it  a toroidal shell of main radius $\rc$ and circular section carrying a uniform surface current generates, at the first order in the $e$-parameter, the same exterior magnetic potential as a circular loop of radius $\rc$ carrying the same current.}\\

The theorem holds in the ${\cal O}(e^2)$ order. It applies likewise to the magnetic field $\vec{B} = \nabla \times \vec{A}$. The $e^2$-approximations for the poloidal components $B_R$ and $B_Z$ of the shell are then given by
\begin{flalign}
    B_R = \frac{\mu_0 I}{2 \pi} \frac{Z}{R\deltao} \left[\frac{R^2+\rc^2+Z^2}{(R-\rc)^2+Z^2} \elie(\ko) - \elik(\ko)\right],
   \label{eq:br}
\end{flalign}
\begin{flalign}
    B_Z = \frac{\mu_0 I}{2 \pi} \frac{1}{\deltao} \left[ - \frac{R^2-\rc^2+Z^2}{(R-\rc)^2+Z^2} \elie(\ko) + \elik(\ko)\right].
   \label{eq:bz}
\end{flalign}

We can deduce the magnetic potential and field of a solid torus carrying a uniform current density $\vec{J}=J_\phi \vec{e_\phi}$, following the procedure given in Sec. \ref{sec:solidtorus}. The $e^2$-approximation for the vector potential is obtained from (\ref{eq:aorder0}) where $S_{0,0}$ is to be replaced by $V_{0,0}$ (which is also unity in the present case). So we can state that (similarity theorem 5):\\

\noindent {\it a toroid of main radius $\rc$ and circular section carrying a uniform volume current density generates, at the first order in the $e$-parameter,, the same exterior magnetic potential (and the field) as a circular loop of radius $\rc$ carrying the same current.}\\

The reader can verify that this theorem also works for a core-stratified current, as for the gravitationnal problem.

  \section{General comments}
\label{sec:comments}

\begin{figure}
  \includegraphics[width=8.7cm,bb=50 50 410 302,clip=true]{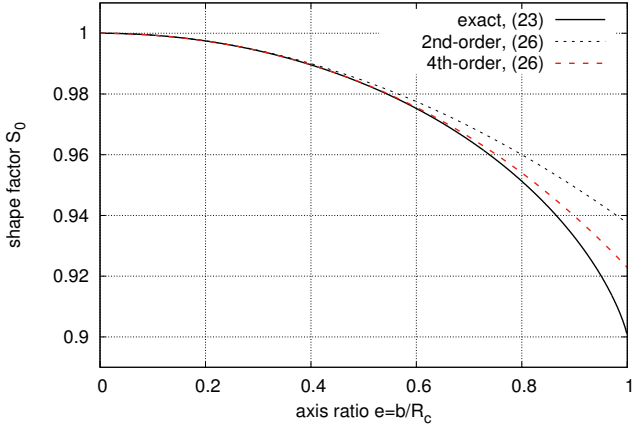}
\caption{The quantity $S_{0,0}$ given by (\ref{eq:s0bis}).}
\label{fig:S00.eps}
\end{figure}

The paper resides on the expansion of $\elik(k)/\Delta$. Other options are possible as quoted before. If we expand $k \elik(k)$ instead of $\elik(k)/\Delta$ in the Green function, one can show that $g_{0,0}$ is changed for
\begin{flalign}
g_{0,0} =  \frac{r}{\Delta_0} \frac{2}{\pi}\elik(\ko) \times S_{0,0},
\label{eq:g00}
\end{flalign}
where
\begin{flalign}
  \label{eq:s0bis}
  S_{0,0}&=\frac{2}{\pi}\elie(p) \sqrt{1+e},
\end{flalign}
\begin{equation}
\elie(k) = \int_0^{\pi/2}{\sqrt{1-k^2 \sin^2\vartheta}d\vartheta}
\end{equation}
is the complete elliptic integral of the second kind, and
\begin{equation}
p^2=\frac{2e}{1+e}\in [0,1].
\label{eq:p}
\end{equation}
With this approach, again, $S_{0,0}$ still does not depend on $R$ and $Z$, but solely on $e$. It is plotted in Fig. \ref{fig:S00.eps}. As we can see, its range of variation, namely $[\frac{2\sqrt{2}}{\pi},1]$, is very small. As a consequence, (\ref{eq:f00}) and (\ref{eq:g00}) are very close, and Fig. \ref{fig:fclose.ps} is almost unchanged. Besides, we have
\begin{flalign}
S_{0,0} = 1- \frac{e^2}{16}- \frac{15e^4}{1024} + \dots
\end{flalign}
for $e \le 1$. Since $S_{0,0}=1$ for $e=0$, we still have $\lim_{e \rightarrow 0} \Psi_0 = \Psi^{\rm loop}$. The similarity theorem 1 reads in this case:\\

  {\it a homogeneous toroidal shell of mass $M$, main radius $\rc$ and circular section of radius $b=e \rc$ generates, at the second order in the $e$-parameter, the same exterior potential as a circular loop of radius $\rc$ and mass $M S_{0,0}$, where $S_{0,0}$ is given by (\ref{eq:s0bis}).}\\

It can be shown after some algebra that the next three surface factors are respectively
\begin{flalign}
  S_{1,0}
  &=\frac{2}{3 \pi}\sqrt{1+e} \left[\elie(p)-(1-e) \elik(p)\right],
\end{flalign}
\begin{flalign}
  S_{2,0}
  &=\frac{8}{15\pi}\sqrt{1+e} \left[\left(\frac{9}{2}e^2-1\right)\elie(p)+(1-e) \elik(p)\right],
\end{flalign}
and
\begin{flalign}
 S_{0,2}
  &=\frac{4}{15 \pi}\sqrt{1+e} \left[(3e^2+1)\elie(p)-(1-e) \elik(p)\right].
\end{flalign}
and the corresponding error map is, as verified, similar to Fig. \ref{fig:psi_err2.ps}.

Another important comment concerns the point where the expansion is performed. In \cite{htkl19}, the choice for the expansion at the focal ring $a=\rf$ was strategical: this is the only point in space which makes the modulus $k$ of the elliptic integral constant all along the circular section of the shell. The motivation for chosing the centre of the circular section here (instead of the focal ring) is similar: the calculation of the integral in (\ref{eq:psishell}) is facilitated, in particular through the expression for $d \ell = b d \theta$. If we use for instance the toroidal coordinates $(\eta, \zeta) \in [0,\infty[ \times [-\pi,\pi[$, the integral over $\theta$ in (\ref{eq:psishell}) can be converted into an integral over $\zeta$. We have in this case
  \begin{flalign}
  a=\rf \frac{\sinh \eta}{\cosh \eta - \cos \zeta},
  \qquad
  z=\rf \frac{\sin \zeta}{\cosh \eta - \cos \zeta},
  \end{flalign}
  where $\rf$ is the radius of the pole (or focal ring), and the line element is $d \ell = \rf \frac{d \zeta}{\cosh \eta - \cos \zeta}$. As a consequence, the potential writes
\begin{flalign}
 \Psi(\vec{r}) = -4G \rf^2 \int_{-\pi}^{\pi} { \frac{\elik(k)}{\Delta}\sinh \eta   \frac{\Sigma(\zeta) d \zeta}{(\cosh \eta - \cos \zeta)^2}},
\label{eq:psishell_focalring}
\end{flalign}
where the modulus $k$ and $\Delta$ depend on $\zeta$. The expansion of $\kappa$ in $x_0=\rf$ (and $y_0=0$ still; see Sects. \ref{sec:expansion} and \ref{sec:gen}) generates, for the homogeneous shell, integrals of the form
\begin{flalign}
\label{eq:integrals_focalring}
  \rf^{n+m+2} \int{\frac{(\cosh \eta - \cos \zeta- \sinh \eta)^{n}}{(\cosh \eta - \cos \zeta)^{n+m+2}}\sin^m \zeta d \zeta}.
\end{flalign}
At order zero (i.e., for $n=m=0$), we find
\begin{flalign}
\Psi(\vec{r}) \approx -8G \rf^2  \Sigma_0  \newkappa_0 S_{0,0}, 
\label{eq:psishell_focalring}
\end{flalign}
where
\begin{flalign}
  S_{0,0} &= 2 \sinh \eta_0 \int_{-\pi}^{\pi}{\frac{ d \zeta}{(\cosh \eta_0 - \cos \zeta)^2}},\\
  \nonumber
  & = \frac{\cosh \eta_0}{\sinh^2 \eta_0} = \frac{b \rc}{\rf^2},
\end{flalign}
and so we recover (\ref{eq:psiorder0final}). For higher terms, (\ref{eq:integrals_focalring}) have to be calculated analytically for all pairs $(n,m)$, and this manifestly requires more effort than for the $J_{nm}$'s.
    
\section{Conclusion and perspectives}
\label{sec:conclusion}

The exterior potential of a static thin toroidal shell, as given by the Laplace equation, is obtained from a double Taylor expansion of the axisymmetric Green function. Each term is then integrated over the source, as in the multipole theory. Here, the expansion is performed at the centre of the circular section instead of the origin of coordinates. The series converges very well and provides a solution which satisfies the Laplace equation in every order, so no ``ghost'' sources are induced by truncation. In practice, the efficiency of the method is remarkable, with already $3$ correct digits at order zero for toroids having an axis ratio of $0.1$. At order $2$, this precision is almost doubled (to $6$ digits), which should be sufficient for most applications.

At order $2$ in the shell parameter (minor-to-major radius ratio), a shellular, solid or core-stratified toroid generates an {\it exterior} potential (and field) similar to that of a thin circular loop having same main radius and same mass. We meet the results by \cite{ba11} and \cite{kondratyev18}. A few similarity theorems, which all resemble the Gauss theorem, have been proposed. The approximations for the exterior potential reported here together with the interior solutions reported in \cite{htkl19} yield a complete decription of the potential of a toroidal shell of circular section, at any point of space. It then becomes possible to deduce the interior solution for the solid torus, since both interior and exterior shell solutions are required in this operation. Next, the energy for the formation of a solid torus becomes accessible. It would be worth to generalize the method to any kind of source shape, not limited to circular section, through specific prescriptions for $a(\theta)$ and $z(\theta)$, or $z(a)$. This would open exciting perspectives, in particular for oblate structures such as geometrically thin discs.

\section*{Acknowledgments}

We are grateful to the referee, Dr M. Majic, for useful comments, and in particular for pointing out the article by \cite{kondratyev18} we were not aware of at the time of submission.

V. Karas acknowledges Czech Science Foundation project No. 19-01137J. A. Trova acknowledges support from the Research Training Group 1620 ``Models of Gravity" funded by the German Science Foundation DFG. O. Semer\'ak is grateful for support from GACR-17/13525S grant of the Czech Science Foundation. We thank V. Tikhonchuk for references about terrestrial plasmas.

\bibliographystyle{mn2e}

\appendix
\onecolumn

\newpage
\section{Integrals $J_{n,m}$}
\label{sec:jnm}

From \cite{gradryz07}, we have
\begin{flalign}
\label{eq:betafunction}
  \int_0^{\pi/2}{ \cos^n \theta \sin^m \theta  d\theta} = \frac{1}{2}B\left(\frac{n+1}{2},\frac{m+1}{2}\right),
\end{flalign}
where $B(x,y)=\frac{\Gamma(x)\Gamma(y)}{\Gamma(x+y)}$ is the complete Beta function, and $\Gamma(x)$ is the Gamma function. From this expression, we can easily deduce $J_{n,m}$ (integral bounds $0$ and $2\pi$). We find 
\begin{flalign}
J_{n,m}& =  \frac{1}{2}B\left(\frac{n+1}{2},\frac{m+1}{2}\right)\left[1+(-1)^n\right]\left[1+(-1)^{n+m}\right].
\end{flalign}
It follows that $J_{n,m}=0$ when $n$ is even or when $n+m$ is even ($n$ and $m$ have different parity). The expression for $J_{n,m}$ are given in Tab. \ref{tab:jnm} for $n=\{0,1,2,3,4\}$ and $m=\{0,1,2,3,4\}$.

\section{Residual mass density}
 \label{eq:residual}

The residual density is found from the Poisson equation, i.e. 
\begin{flalign}
  \nabla^2 \left[\frac{\elik(k)}{\sqrt{(a+R)^2+Z^2}} \right] =  \frac{\ko\elik(\ko)}{4R^2} + \left[\frac{\partial^2 k}{\partial R^2}+\frac{\partial k^2}{\partial Z^2} \right] \frac{\elie(\ko)}{\kpo^2} +  \left[\left(\frac{\partial k}{\partial R}\right)^2+\left(\frac{\partial k}{\partial Z}\right)^2 \right] \frac{(1+\ko^2)\elie(\ko)-\kpo^2 \elik(\ko)}{k \kpo^4},
\end{flalign}
where the partial derivatives of $\elik(k)$ and  $\elie(k)$ with respect to the modulus $k$ are found in mathematical textbooks \citep{gradryz07}. By expanding all the terms inside the curly brackets, this quantity is strictly zero provided $a-R \ne 0$ and $Z \ne 0$, which never occurs in free space.

\section{Partial derivatives}
\label{sec:derivatives_kk}

There are different ways to calculate the partial derivatives of $\newkappa$ with respect to $a$ and $z$. We find convenient to rewrite $\elik(k)$ as the definite integral over the azimuth, i.e. (\ref{eq:elik}). The denominator is then expanded and rearranged so that the $n$-order derivative with respect to $a$ and $z$ writes
\begin{flalign}
   \frac{\partial^n \newkappa}{\partial a^{n-m} \partial z^m} & = \frac{\partial^n}{\partial a^{n-m} \partial z^m} \int_0^{\frac{\pi}{2}}{\frac{d \phi}{\sqrt{\Delta^2 - 4 aR \sin^2 \phi}}}\\
  \nonumber
 & = \int_0^{\frac{\pi}{2}}{d \phi\frac{\partial^n}{\partial a^{n-m} \partial z^m} \left\{\left[a+R \cos(2\phi)\right]^2+\left[R\sin(2\phi)\right]^2 + \zeta^2 \right\}^{-1/2}}.
\end{flalign}
Denoting $D=\left[a+R \cos(2\phi)\right]^2+\left[R\sin(2\phi)\right]^2 + \zeta^2$, we have
\begin{flalign}
   \frac{\partial D^{-1/2}}{\partial a} = -\left[a+R\cos(2\phi)\right] D^{-3/2}
\end{flalign}
and
\begin{flalign}
   \frac{\partial^2 D^{-1/2}}{\partial a^2} = -D^{-3/2}+3\left[a+R\cos(2\phi)\right]^2D^{-5/2}.
\end{flalign}
It follows that
\begin{flalign}
  \frac{\partial \newkappa}{\partial a} &=  \int_0^{\frac{\pi}{2}}{\frac{\partial D^{-1/2}}{\partial a} d\phi} = -(a+R) \Delta^{-3} \frac{\elie(k)}{\kp^2} +2R  \Delta^{-3} \frac{\elie(k)-\kp^2 \elik(k)}{k^2\kp^2}
\end{flalign}
and
\begin{flalign}
  \frac{\partial^2 \newkappa}{\partial a^2}  &=  \int_0^{\frac{\pi}{2}}{\frac{\partial^2 D^{-1/2}}{\partial a^2} d\phi}\\
  \nonumber
  &= -\Delta^{-3} \frac{\elie(k)}{\kp^2}+3\Delta^{-3} \frac{\elie(k)}{\kp^2}-3\zeta^2\Delta^{-5} \frac{2(1+\kp^2)\elie(k)-\kp^2 \elik(k)}{3\kp^4} - 12R^2\Delta^{-5}\frac{(1+\kp^2)\elie(k)-2\kp^2 \elik(k)}{3k^4\kp^2},\\
    &= 2\Delta^{-3} \frac{\elie(k)}{\kp^2}-\zeta^2\Delta^{-5} \frac{2(1+\kp^2)\elie(k)-\kp^2 \elik(k)}{\kp^4} - 4R^2\Delta^{-5}\frac{(1+\kp^2)\elie(k)-2\kp^2 \elik(k)}{k^4\kp^2},
  \nonumber
\end{flalign}

For the $z$-derivatives, we have
\begin{flalign}
   \frac{\partial D^{-1/2}}{\partial z} = \zeta D^{-3/2}, \qquad    \frac{\partial^2 D^{-1/2}}{\partial z^2} = -D^{-3/2}+3\zeta^2 D^{-5/2}
\end{flalign}
and, consequently,
\begin{flalign}
  \frac{\partial \newkappa}{\partial z} & = \zeta \int_0^{\frac{\pi}{2}}{\frac{\partial D^{-1/2}}{\partial z} d\phi} = \zeta \Delta^{-3} \frac{\elie(k)}{\kp^2}
\end{flalign}
and
\begin{flalign}
  \frac{\partial^2 \newkappa}{\partial z^2} &=  \int_0^{\frac{\pi}{2}}{\frac{\partial^2 D^{-1/2}}{\partial z^2} d\phi}= -\Delta^{-3} \frac{\elie(k)}{\kp^2}+ \zeta^2 \Delta^{-5} \frac{2(1+\kp^2)\elie(k)-\kp^2 \elik(k)}{\kp^4}.
\end{flalign}


\section{F90 program for the exterior potential}
\label{sec:f90program}

\begin{verbatim}
Program F90drivercode3
  ! "The exterior gravitational potential of toroids"
  ! Hure, Basillais, Karas, Trova & Semerak (2019), MNRAS
  ! gfortran F90drivercode3.f90; ./a.out
  ! not optimized
  Implicit None
  Integer,Parameter::AP=Kind(1.00D+00)
  Real(Kind=AP),Parameter::PI=ATAN(1._AP)*4
  Real(KIND=AP)::B,RC,MASS,E ! core radius, main radius and mass of the shell, and axis ratio
  Real(KIND=AP)::KMOD,KMOD2,KPRIM,KPRIM2 ! moduli
  Real(KIND=AP)::R,Z,PSI ! cylindrical coordinates and potential value where it is estimated
  Real(KIND=AP)::Z2,VAL,DELTA,DELTA2,DELTA3,DELTA4,DELTA5 ! misc
  Real(KIND=AP)::KMOD4,KPRIM4,S0,S10,S20,S02,D32,S2D32,D52,S2C2D52 ! misc
  Real(KIND=AP)::ELLIPTICK,ELLIPTICE ! complete elliptic integrals
  ! ? input parameters (properties of the shell)
  B=0.1_AP
  RC=1._AP
  E=B/RC
  MASS=B*RC*PI**2*4
  print*,"Mass of the shell",MASS
  ! ? values of R and Z where the potential is requested (must be outside the cavity!)
  R=RC*2
  Z=RC*2
  Z2=Z**2
  If ((R-RC)**2+Z2-B**2<0._AP) Then
     ! approximation not valid inside the shell
     PSI=0._AP
  Else
     DELTA2=(R+RC)**2+Z2
     KMOD2=RC*R*4/DELTA2
     DELTA=Sqrt(DELTA2)
     KPRIM2=((R-RC)**2+Z2)/DELTA2
     ! values of K(k) and E(k) to be set here !
     ! ELLIPTICE=
     ! ELLIPTICK=
     !misc.
     DELTA3=DELTA2*DELTA
     DELTA5=DELTA3*DELTA2
     KMOD4=KMOD2**2
     KPRIM4=KPRIM2**2
     ! surface factors
     S0=1._AP
     S10=E**2/2
     S20=E**2/2
     S02=S20
     ! coefficients
     D32=ELLIPTICE/KPRIM2
     S2D32=(ELLIPTICE-KPRIM2*ELLIPTICK)/KPRIM2/KMOD2
     D52=((1._AP+KPRIM2)*ELLIPTICE*2-KPRIM2*ELLIPTICK)/KPRIM4/3
     S2C2D52=((1._AP+KPRIM2)/KPRIM2*ELLIPTICE-ELLIPTICK*2)/KMOD4/3
     ! order 0
     VAL=ELLIPTICK/DELTA*S0
     ! order 1
     VAL=VAL+RC*(-(R+RC)*D32/DELTA3+S2D32/DELTA3*R*2)*S10
     ! order 2
     VAL=VAL+RC**2*((-D32/DELTA3+D32/DELTA3*3-D52*Z2/DELTA5*3-S2C2D52/DELTA5*R**2*12)*S20&
          &+(-D32/DELTA3+D52*Z2/DELTA5*3)*S02)/2
     PSI=-VAL*B*RC*PI*8
     Print *,"Potential value (2nd-order)",PSI,PSI/MASS*RC
  Endif
End Program F90drivercode3
\end{verbatim}

\end{document}